\documentclass[3p,times,preprint]{elsarticle}

\usepackage{amssymb}
\usepackage{float}
\usepackage{subfig}
\usepackage{amsmath}

\usepackage[figuresright]{rotating}

\begin{document}

\begin{frontmatter}




\title{A cancellation problem in hybrid particle-in-cell schemes due to finite particle size}


\author{A.~Stanier, L.~Chac\'{o}n, A.~Le}

\address{Applied Mathematics and Plasma Physics, Los Alamos National Laboratory, Los Alamos, NM 87545, United States}

\begin{abstract}
The quasi-neutral hybrid particle-in-cell algorithm with kinetic ions and fluid electrons is a popular model to study multi-scale problems in laboratory, space, and astrophysical plasmas. Here, it is shown that the treatment of ions as finite-size particles and electrons as a grid-based fluid can cause significant numerical wave dispersion errors in the magnetohydrodynamic limit ($kd_i \ll 1$, where $d_i$ is the ion skin-depth). Practical requirements on the mesh spacing $\Delta x/d_i$ are suggested to bound these errors from above. 
\end{abstract}


\end{frontmatter}


\section{Introduction}\label{}

Particle-in-cell (PIC) methods~\cite{birdsall91,hockney81} are widely used to model kinetic plasma physics problems as they avoid the need to solve for the plasma distribution function on a 6D (3D-3V) grid, and they can be highly optimized to run on modern computer architectures with multiple levels of parallelism~\cite{bowers09}. However, care must be taken as PIC simulations can potentially suffer from a number of algorithmic issues that are not commonly found in purely grid-based codes. Issues relate to statistical noise from the use of a finite number of macro-particles~\cite{nevins05}, and the numerical heating of these particles due to lack of discrete conservation properties~\cite{birdsall91,rambo95}. To partially mitigate such effects, macro-particles are given finite spatial size to smooth the particle-grid interaction and grid based filtering can be applied to hydrodynamic moments and electromagnetic fields~\cite{birdsall91}. These techniques can also cause unwanted attenuation of the physical signal of interest at lower wavenumber $k$, but such errors can be made small if the problem is well resolved ($k\Delta x \ll 1$). 

The hybrid-PIC scheme differs from the fully kinetic PIC method in that the electrons are treated as a grid-based fluid~\cite{byers78,hewett78,winske03,stanier19cart}. This is done to enable the study of problems in which the coupling between macroscopic and ion kinetic scales is important~\cite{winske85,chapman15,karimabadi14,stanier15prl,le18}, without the need to resolve stiff electron scales. However, algorithmic limitations in the hybrid-PIC approach have been less well studied than for fully kinetic PIC. In this note, it is shown that errors from the use of finite sized particles and/or smoothing can potentially become large for the hybrid-PIC model, even for the case with $k\Delta x \ll 1$. To obtain the correct long wavelength magnetohydrodynamic limit the electric field term must cancel when taking the sum of the ion and electron momentum equations. However, as described below, this does not occur exactly in hybrid-PIC due to the different spatial discretization of ions and electrons.

\section{Hybrid-PIC algorithm}
\subsection{Semi-discrete formulation}

The cold plasma kinetic-ion and fluid-electron hybrid model is considered in linearized and semi-discrete form for transverse electromagnetic waves propagating parallel to a background magnetic field $\boldsymbol{B}_0 = B_0 \boldsymbol{\hat{x}}$. To solve the Vlasov equation, $d_t f_i = 0$, the ion distribution function is sampled by macro-particle markers as $f_i(t,\boldsymbol{x},\boldsymbol{v}) \approx \sum_p S_m(x - x_p) \delta(\boldsymbol{v}-\boldsymbol{v}_p(t))$. Here, the finite-size particle shape functions $S_m$ are $m$-th order B-splines, which have compact support and form a partition of unity. The markers are advanced as
\begin{equation}m\frac{d (\boldsymbol{\delta v}_p)}{dt} = e\left(\boldsymbol{\delta E}_p + \boldsymbol{\delta v}_p \times \boldsymbol{B}_0\right), \quad \quad \delta x_p = 0,\end{equation}
where $e$ and $m$ are the ion particle charge and mass. The particle positions are stationary along the $x$-direction due to the cold plasma assumption with only transverse electric fields.

Grid-based quantities $\chi_g$ are defined at cell centers, and derivatives are computed using second order finite  differences. The electric field in the non-relativistic quasi-neutral limit is calculated from Ohm's law as
\begin{equation}\label{gridohms}\boldsymbol{\delta E}_g = - \boldsymbol{\delta u}_g \times \boldsymbol{B}_0 + \frac{\left(\boldsymbol{\nabla}_g \times \boldsymbol{\delta B}_g\right) \times \boldsymbol{B}_0}{\mu_0e n_0},\end{equation}
where $\mu_0$ is the magnetic constant and $n_0$ is the background density. Faraday's equation is used to advance the magnetic field
\begin{equation}\label{faradayohms}\frac{\partial \boldsymbol{\delta B}_g}{\partial t} = - \boldsymbol{\nabla}_g \times \boldsymbol{\delta E}_g.\end{equation}

To close the system, the perturbed ion velocity moment is gathered from the particles to the grid as
\begin{equation}\boldsymbol{\delta u}_g  = \frac{1}{n_0\Delta x}\textrm{SM}_g\left(\sum_p S_m(x_g - x_p) \boldsymbol{\delta v}_{p}\right),\end{equation}
where $\Delta x$ is the cell size, and the electric field is scattered from the grid to the particle positions as
\begin{equation}\boldsymbol{\delta E}_p = \sum_g S_m(x_g - x_p) \textrm{SM}_g\left(\boldsymbol{\delta E}_g\right).\end{equation}
Here $\textrm{SM}_g$ is an optional binomial smoothing operator that acts on grid quantities to reduce noise. It is defined as $\textrm{SM}_g(\chi_g) = (\chi_{g-1} + 2 \chi_{g} + \chi_{g+1})/4$.

\subsection{Semi-discrete dispersion relation}

To derive a semi-discrete dispersion relation, it is assumed the number of particles is large such that an ion momentum equation can be defined in the continuum. Taking the finite-domain, continuous Fourier transform of this momentum equation gives
\begin{equation}\label{fouriermomentum}-i\omega m n_0 \widetilde{\boldsymbol{\delta u}} = en_0 \left[\widetilde{\boldsymbol{\delta E}} + \widetilde{\boldsymbol{\delta u}} \times \boldsymbol{B}_0\right],\end{equation}
where $\widetilde{\delta \chi}$ are the Fourier mode amplitudes of the continuum space variables. 

Eqs.~(\ref{gridohms}-\ref{faradayohms}) are defined on a spatial grid. Using the finite-domain, discrete Fourier transform gives 
\begin{equation}\label{fourierohms}\widetilde{\boldsymbol{\delta E}_g} = -\widetilde{\boldsymbol{\delta u}_g} \times \boldsymbol{B}_0 + \frac{(i \boldsymbol{\kappa} \times \widetilde{\boldsymbol{\delta B}_g)} \times \boldsymbol{B}_0}{\mu_0 e n_0},\end{equation}
\begin{equation}\label{fourierfaraday}-i\omega \widetilde{\boldsymbol{\delta B}_g} = - i \boldsymbol{\kappa} \times \widetilde{\boldsymbol{\delta E}_g},\end{equation}
where $\boldsymbol{\kappa} = \boldsymbol{\hat{x}}[k \sin{(k\Delta x)}]/(k\Delta x)$ is the modification to the wavenumber from the finite-difference approximation. 

The transformed continuum electric field relates to the transformed discrete (grid) electric field as
\begin{equation}\label{fourierEp}\widetilde{\boldsymbol{\delta E}} = \textrm{SM}(k\Delta x)S_m(-k\Delta x)\widetilde{\boldsymbol{\delta E}_g},\end{equation}
where $\textrm{SM}(k\Delta x) = \cos^2{(k\Delta x/2)}$. The transformed discrete ion velocity moment relates to the transformed continuum moment as (e.g.~\cite{birdsall91})
\begin{equation}\label{fourierUg}\widetilde{\boldsymbol{\delta u}_g} = \textrm{SM}(k\Delta x)\sum_qS_m(k_q\Delta x)\widetilde{\boldsymbol{\delta u}}(k_q),\end{equation}
where the sum is taken over the aliases $q\in \mathbb{Z}$ where $k_q = k - 2\pi q/\Delta x$. Eq.~(\ref{fouriermomentum}) can be written in terms of transformed discrete quantities using Eqs.~(\ref{fourierEp}-\ref{fourierUg}), as
\begin{equation}\label{fouriermomentumdiscrete}-i\omega m n_0 \widetilde{\boldsymbol{\delta u}_g} = en_0 \left |\textrm{SM}(k\Delta x)\right |^2 \sum_q\left|S_m(k_q\Delta x)\right|^2\widetilde{\boldsymbol{\delta E}_g} + en_0\widetilde{\boldsymbol{\delta u}_g} \times \boldsymbol{B}_0,\end{equation}
where the periodicity property has been used for $\textrm{SM}(k_q\Delta x)=\textrm{SM}(k\Delta x)$ and $\widetilde{\boldsymbol{\delta E}_g}(k_q)=\widetilde{\boldsymbol{\delta E}_g}(k)$. 

\section{Hybrid cancellation problem}\label{}

The resulting dispersion relation is found from Eqs.~(\ref{fourierohms}, \ref{fourierfaraday}, \ref{fouriermomentumdiscrete}) as
\begin{equation}\label{dispersioneq}\omega = \pm v_A \kappa \left(\sqrt{1+\frac{1}{4}\left[d_i \kappa - \frac{1 - \left| SM(k\Delta x)\right|^2 \sum_q \left|S_m(k_q\Delta x)\right|^2 }{d_i \kappa}\right]^2} \pm \frac{1}{2} \left[d_i \kappa - \frac{1-\left| SM(k\Delta x)\right|^2 \sum_q \left| S_m(k_q\Delta x)\right|^2}{d_i\kappa}\right]\right),\end{equation}
where $d_i = v_A/\Omega_{ci}$ is the ion skin-depth, $v_A=B_0/\sqrt{\mu_0 n_0 m}$ is the Alfv\'en velocity, and $\Omega_{ci}=eB_0/m$ is the ion cyclotron frequency. It is instructive to compare this semi-discrete dispersion relation with the physical result ($\Delta x \rightarrow 0$), given by
\begin{equation}\label{continuumalfvenwhistler}\omega_{\textrm{ph}} = \pm v_A k \left(\sqrt{1 + \frac{1}{4} d_i^2 k^2} \pm \frac{1}{2} d_i k\right).\end{equation}

In addition to the standard finite-difference modification of the wavenumber $k\rightarrow \kappa(k)$, there are additional unphysical terms resulting from the Fourier representations of the shape functions and the smoothing operators. At this stage, the hybrid cancellation problem can be discerned: the presence of $d_i \kappa$ in the denominator of these unphysical terms may cause them to become arbitrarily large as $d_i k\rightarrow 0$.

\begin{figure}
\begin{center}
\subfloat[Right-hand Alfv\'en-whistler]{\includegraphics[trim={3cm 8cm 4cm 7.6cm},clip,width=0.48\textwidth]{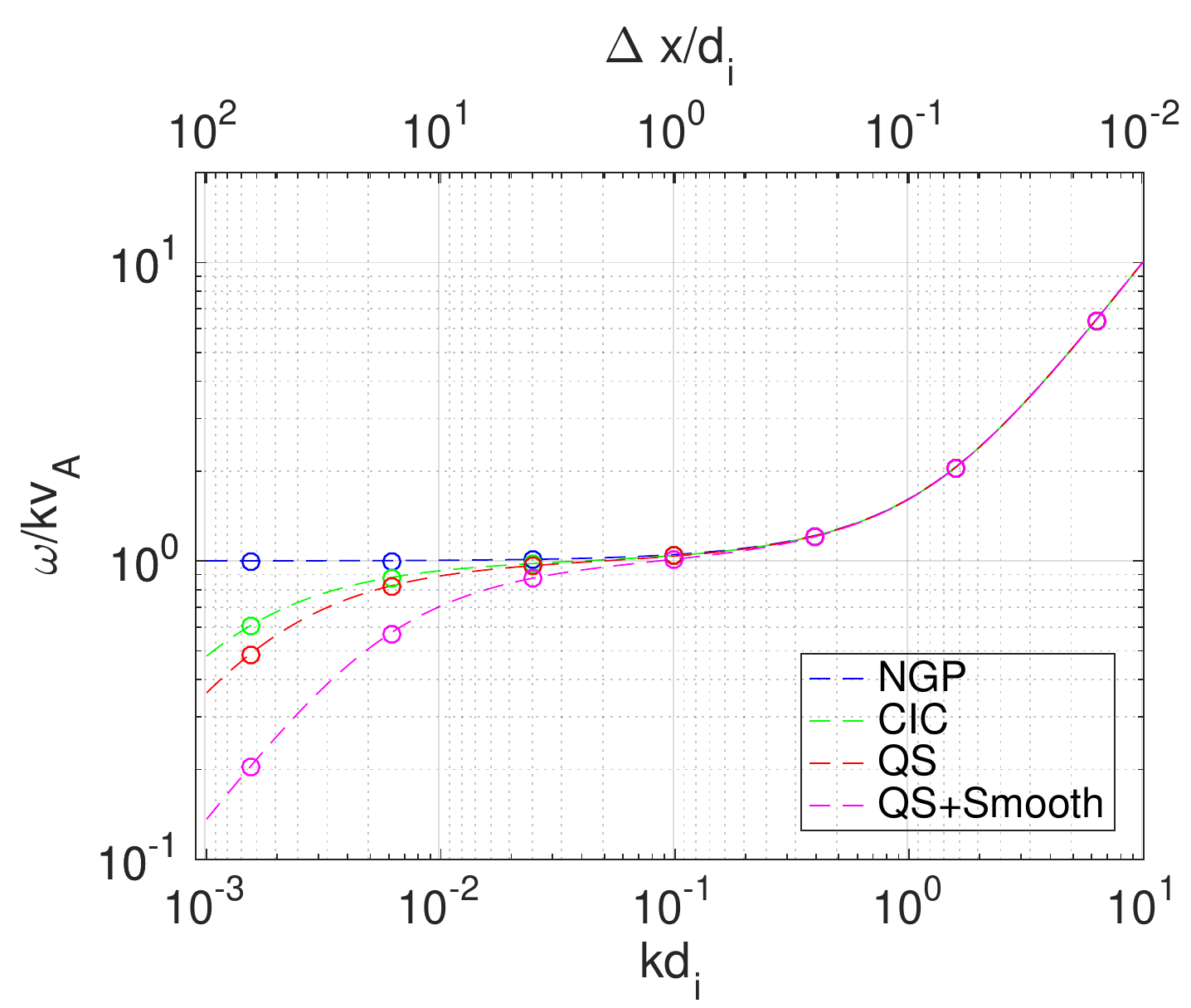}}
\subfloat[Left-hand Alfv\'en-cyclotron]{\includegraphics[trim={3cm 8cm 4cm 7.6cm},clip,width=0.48\textwidth]{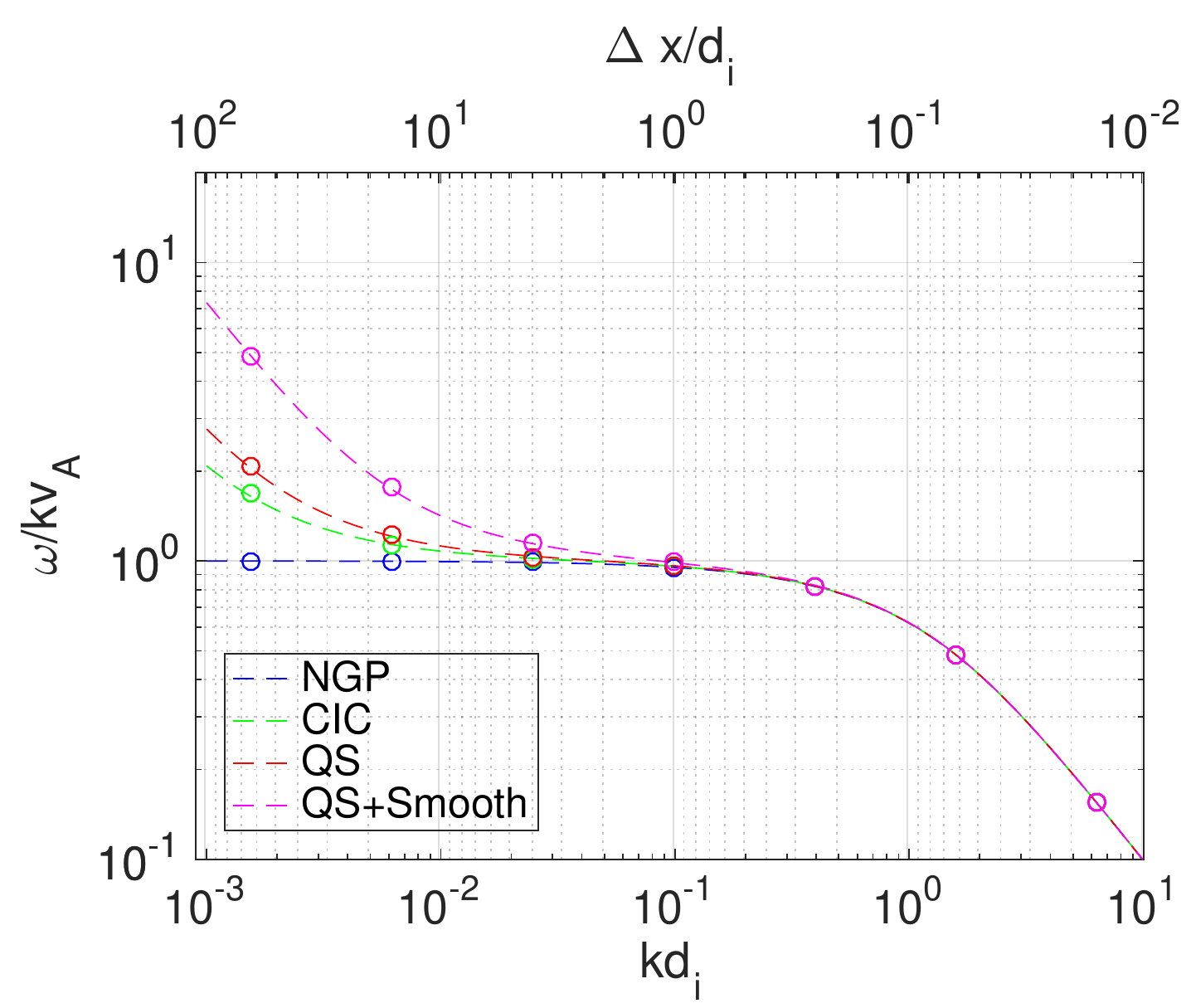}}
\caption{\label{dispersionplot}Semi-discrete dispersion relation from Eq.~(\ref{dispersioneq}) shown as dashed curves for zeroth-order Nearest Grid Point (NGP, blue), first-order Cloud-In-Cell (CIC, green), second-order Quadratic-Spline (QS, red) without smoothing, and Quadratic-Spline with Binomial smoothing (QS+Smooth, magenta) applied to fields and moments. Circles are phase velocities measured from 1D electromagnetic hybrid simulations. Here the wave is well resolved with fixed resolution $k\Delta x=\pi/32$ in each case, while $k d_i$ is varied (bottom axis). The resolution $\Delta x/d_i$ (top axis) varies inversely proportional to $kd_i$ in this case. Only NGP recovers the correct limits $\omega/kv_A \rightarrow 1$ as $kd_i\rightarrow 0$.  }
\end{center}
\end{figure}

To quantify these errors, it is necessary to compute the shape function terms $\sum_q \left| S_m (k_q \Delta x)\right|^2$, which involves analytically calculating the sum over aliases. Following Ref.~\cite{birdsall91}, 
\begin{equation}\label{NGPshapeterm}\textrm{Nearest Grid Point}\,(\textrm{NGP},\,\,m=0): \quad \quad \sum_q \left| S_0 (k_q \Delta x)\right|^2 = 1,\end{equation}
\begin{equation}\label{CICshapeterm}\textrm{Cloud In Cell}\,(\textrm{CIC},\,\,m=1): \quad \quad \sum_q \left| S_1 (k_q \Delta x)\right|^2 = \frac{1}{3}\left[1+2\cos^2\left(\frac{1}{2}k\Delta x\right)\right],\end{equation}
\begin{equation}\label{QSshapeterm}\textrm{Quadratic Spline}\,(\textrm{QS},\,\,m=2): \quad \quad \sum_q \left| S_2 (k_q \Delta x)\right|^2 = \frac{1}{15}\left[2+11\cos^2\left(\frac{1}{2}k\Delta x\right)+2\cos^4\left(\frac{1}{2}k\Delta x\right)\right].\end{equation}

The predicted dispersion relation from Eq.~(\ref{dispersioneq}) is plotted as dashed lines in Fig.~\ref{dispersionplot} for both left and right-hand polarized waves for the cases of NGP (blue), CIC (green), QS (red) without smoothing, and the case of QS with one pass of binomial smoothing (magenta) applied symmetrically to the field and moment quanities.  The overplotted circles show the measured phase velocities from corresponding simulations using a 1D explicit electromagnetic hybrid algorithm, which verify the analytic result. Here, a small time-step is used to give negligible temporal truncation error and the wavelength of the perturbation is well resolved with $64$ grid cells in each case, such that the spatial truncation errors are fixed ($k\Delta x = \pi/32$). The top horizonal axis gives the absolute size of the spatial cells in terms of the ion skin-depth, where $\Delta x/d_i \propto 1/kd_i$ for fixed $k\Delta x$.
 
For the short-wavelength limit ($d_i k \gg 1$), good agreement is found with Eq.~(\ref{continuumalfvenwhistler}) in each case for the right-hand polarized whistler ($\omega \propto k^2$) and the left-hand polarized ion cyclotron wave ($\omega \rightarrow \Omega_{ci}$). However, the correct long-wavelength limit ($\omega/kv_A \rightarrow 1$ as $kd_i \rightarrow 0$) is only recovered for the case of NGP without smoothing, for which the numerator is exactly zero for the unphysical terms in Eq.~(\ref{dispersioneq}). For higher order shape functions, the phase-speed of the right (left) hand polarized waves is reduced (increased). This error increases as the width of the particle shape function is increased, and is further increased by the application of smoothing. The incorrect MHD-limit can be reached due to the inexact cancellation of the electric fields when combining the ion~(\ref{fouriermomentumdiscrete}) and electron~(\ref{fourierohms}) momentum equations to find a total momentum equation, which is due to the convolutional smoothing of the shape function and grid smoothing operators in Eq.~(\ref{fouriermomentumdiscrete}). The hybrid-PIC scheme is only asymptotic preserving~(e.g.~\cite{degond17}) in the spatial sense for NGP. 

\section{Discussion}

For linear problems, it is useful to estimate how large a value of $(\Delta x/d_i)$ can be taken for a given desired accuracy. To second order in the assumed small parameter $(k\Delta x) \ll 1$, $\kappa\approx k[1-(k\Delta x)^2/6]$, and 
\begin{equation}\frac{1 - \left| SM(k\Delta x)\right|^2 \sum_q \left|S_m(k_q\Delta x)\right|^2}{d_i \kappa} \approx C (k\Delta x)\left(\frac{\Delta x}{d_i}\right),\end{equation}
where the constant $C$ depends on the order of shape function and amount of smoothing: $C=0$ for NGP, $C=1/6$ for CIC, $C=1/4$ for QS, and $C=3/4$ for QS with one pass of smoothing to the fields and moments.

The relative dispersion error due to the second order finite-difference approximation, $\epsilon_{\textrm{FD}} = |\omega - \omega_{ph}|/\omega_{ph}$, can be computed by assuming $C=0$. For $kd_i \ll 1$, $\epsilon_{FD} \approx (k\Delta x)^2/6$. This can be compared with the estimated dispersion error contribution solely from the cancellation problem, $\epsilon_{\textrm{CP}}$. Assuming $C\neq 0$, and then taking $(k\Delta x)^2 \ll (k\Delta x)(\Delta x/d_i) \sim \mathcal{O}(1)$, gives $\epsilon_{CP} \approx C(k\Delta x)(\Delta x/d_i)/2$. The cancellation error dominates the finite-difference error and determines the resolution requirements for $kd_i \ll 1$. The minimum mesh-spacing requirement to achieve a desired error $\epsilon$ for a specific wavenumber $(kd_i) \ll 1$ is therefore given by
\begin{equation}\label{resolutionrequirements}\frac{\Delta x}{d_i} \lesssim \sqrt{\frac{2 \epsilon}{C(kd_i)}}.\end{equation}

\begin{figure}
\begin{center}
\subfloat[Nearest Grid Point (NGP). $\Delta x/d_i=1$.]{\includegraphics[trim={2cm 0.4cm 2.5cm 2.5cm},clip,width=0.48\textwidth]{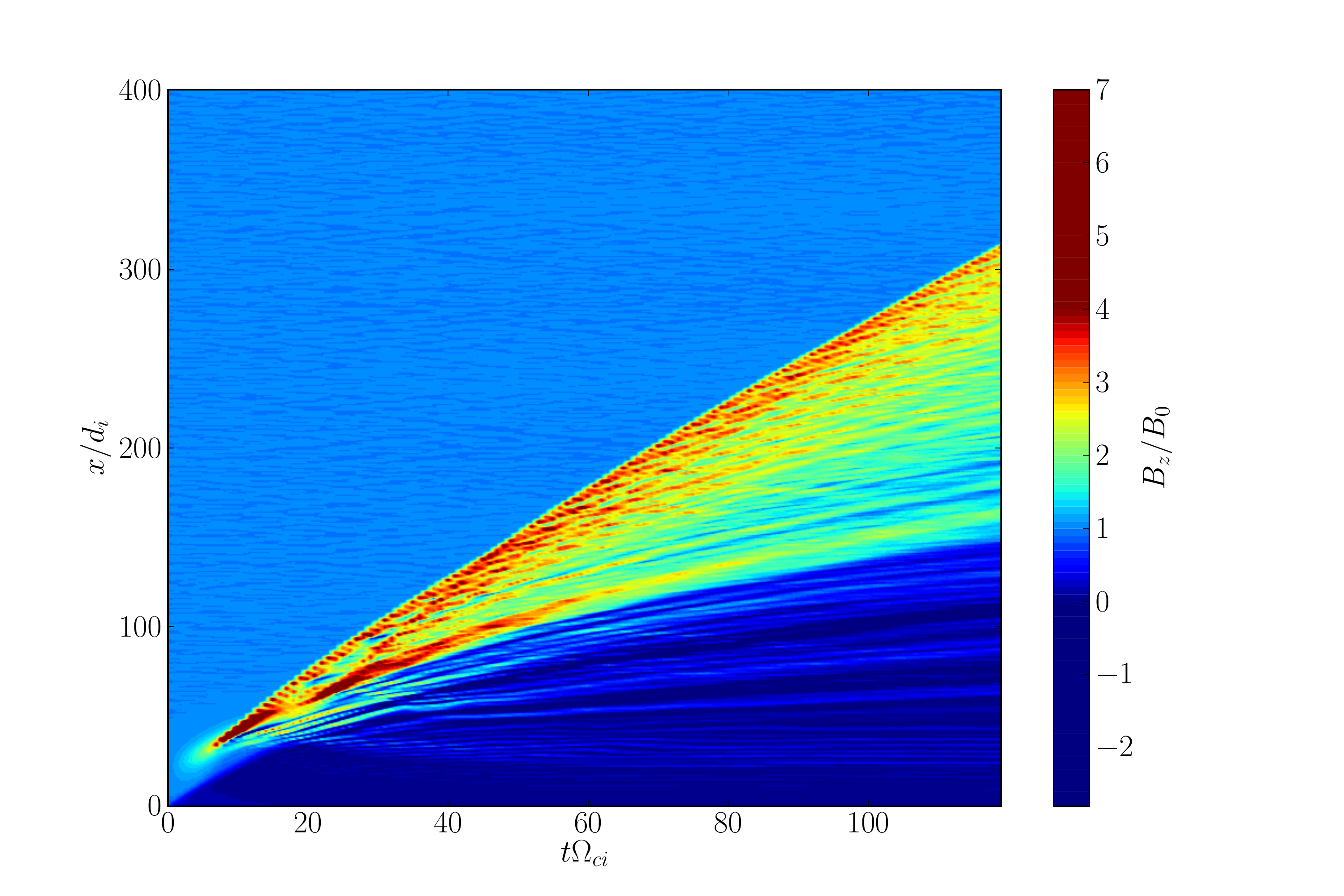}}
\subfloat[Quadratic Spline (QS) with smoothing. $\Delta x/d_i=1$.]{\includegraphics[trim={2cm 0.4cm 2.5cm 2.5cm},clip,width=0.48\textwidth]{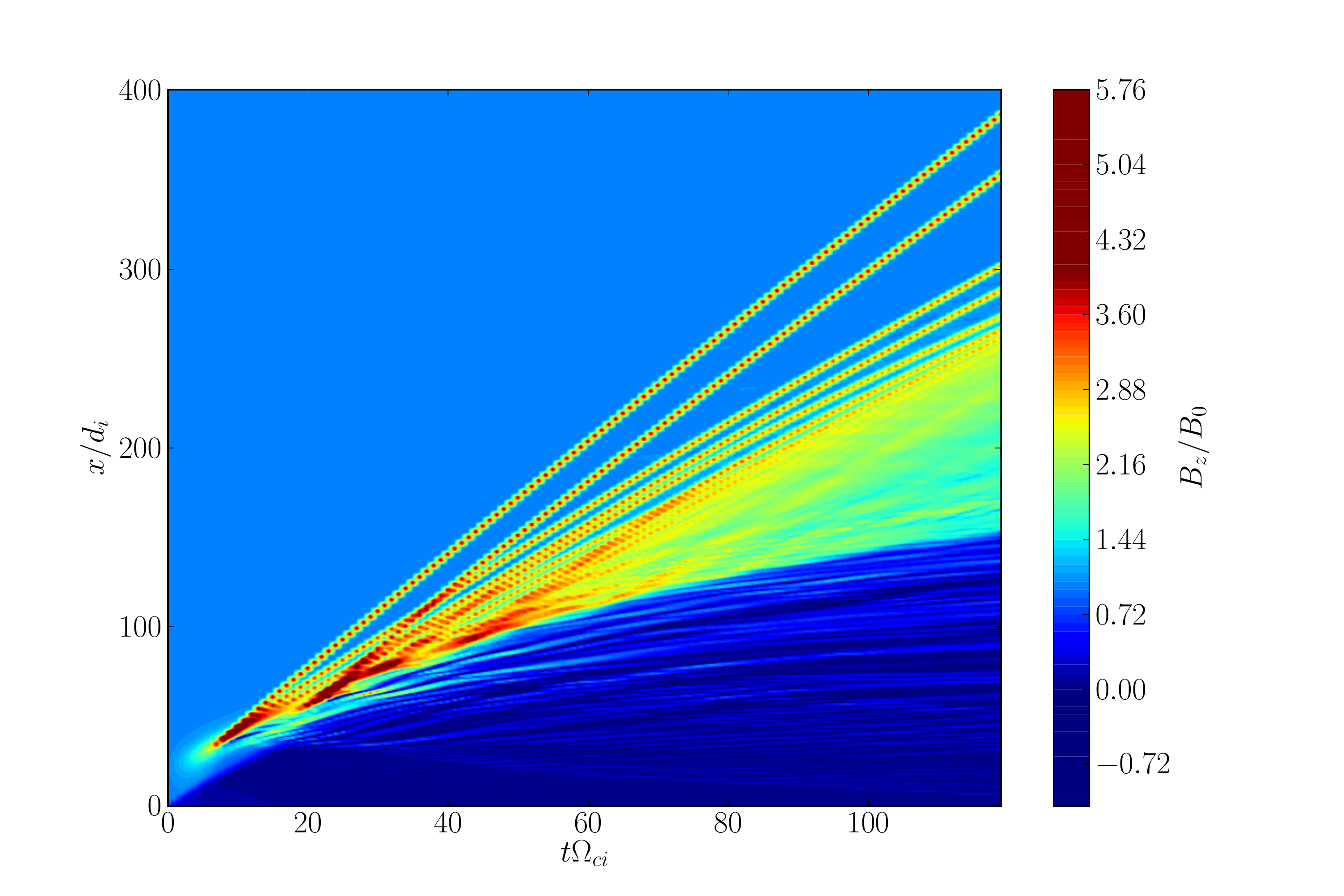}}\\
\subfloat[Quadratic Spline (QS) with smoothing. $\Delta x/d_i=0.5$]{\includegraphics[trim={2cm 0.4cm 2.5cm 2.5cm},clip,width=0.48\textwidth]{bz-exponential-density-QS-WSMOOTH-nx1024-raster.pdf}}
\caption{\label{hane1dtest}a) Simulation with resolution $\Delta x/d_i =1$ using NGP shows formation of perpendicular shock and magnetic cavity caused by the super-Alfv\'enic expansion of debris ions into a uniform magnetized background plasma. b) The same simulation, but with QS shape function and smoothing, gives unphysical solitons due to the interplay of numerical dispersion errors and non-linear steepening. c) At higher resolution $\Delta x/d_i=0.5$, these solitons are reduced (although not completely removed at this resolution). }
\end{center}
\end{figure}

While the above results have been derived for parallel propagating waves with a uniform background density, similar dispersion errors due to the cancellation problem can be found for the case of fast magnetosonic waves propagating perpendicular to a background magnetic field. In Figure~\ref{hane1dtest}, we give a dramatic non-linear numerical example of how such dispersion errors can lead to incorrect physics results. For this simulation, a cloud of debris ions with number density $n_d=(20 n_b/\sqrt{\pi})\exp{(-x^2/(15 d_i)^2)}$ and velocity $v_d = 5 v_{Ab}\boldsymbol{\hat{x}}$ is released into a uniform background plasma with magnetic field $\boldsymbol{B}_0 = B_0 \boldsymbol{\hat{z}}$, density $n_b$, Alfv\'en speed $v_b = B_0/\sqrt{m_b n_b\mu_0}$, cyclotron frequency $\Omega_{ci}=q_b B_0/m_b$ and skin-depth $d_i=v_{Ab}/\Omega_{ci}$.  The ratio of debris ion charge and mass to background values is $q_d/q_b=1$ and $m_d/m_b=3$ respectively. The super-Alfv\'enic expansion of the debris ions excludes the background magnetic field to create a magnetic cavity, and couples with the background ions to create a perpendicular fast magnetosonic shock~\cite{winskegary07}. For higher order shape functions, we observe that numerical dispersion errors are able to support the formation of unphysical solitons that are generated during non-linear steepening when the shock is formed. When followed for long time-scales, these unphysical solitons can detatch and move ahead of the shock wave. Using either NGP shape function, or by sufficiently decreasing $\Delta x/d_i \ll 1$, can remove these artifacts.

The form of cancellation errors in the dispersion relation of Eq.~(\ref{dispersioneq}) appear similar to the cancellation problem found in electromagnetic gyrokinetic algorithms (see e.g.~\cite{cummings96,mishchenko17,mandell19}). However, it is worth noting two differences. Firstly, the cancellation problem in hybrid-PIC is less restrictive than that in gyrokinetics, as it causes dispersion errors at the ion skin-depth scale rather than the electron skin-depth ($d_i/d_e = \sqrt{m_i/m_e} \gg 1$). Secondly, the cancellation problem occurs in gyrokinetics due to the choice of the parallel canonical momentum, $p_\parallel$, as a dependent variable, rather than $v_\parallel$. The $p_\parallel$ formulation is typically chosen for semi-implicit gyrokinetic schemes as the $v_\parallel$ formulation contains an implicit coupling. In fact, the gyrokinetic cancellation problem can be avoided entirely by solving the $v_\parallel$ formulation implicitly~\cite{sturdevant19dpp}. The hybrid-PIC cancellation problem is due to the different spatial discretization of ions (particles) and electrons (grid-based) and does not depend on the choice of time integration scheme.

\section*{Acknowledgements}
A.S. thanks Dan Winske and David Burgess for useful discussions. This material is based upon work supported by the U.S. Department of Energy, Office of Science, Office of Applied Scientific Computing Research (ASCR). This research used resources provided by the Los Alamos National Laboratory Institutional Computing Program, which is supported by the U.S. Department of Energy National Nuclear Security Administration under Contract No. 89233218CNA000001. AL was supported by the Laboratory Directed Research and Development program of Los Alamos National Laboratory under project number 20200334ER.




\bibliographystyle{elsarticle-num}







\end{document}